\documentclass{moriond}
\usepackage{amsmath}
\usepackage{amssymb}

\def\be{\begin{equation}}
\def\ee{\end{equation}}

\newcommand{\Photo}{}

\begin{document}
\vspace*{4cm}
\title{Light-Ion Collisions: Bridging Small and Large QCD Systems}

\author{Aleksas Mazeliauskas}

\address{Institute for Theoretical Physics, Heidelberg University,\\
Philosophenweg 16, 69120 Heidelberg, Germany}

\maketitle\abstracts{
Light-ion collisions at the LHC bridge the gap between small proton-proton and large heavy-ion collision systems, providing a unique laboratory to study the onset of QCD collective phenomena.  The first light-ion run at the LHC took place July~1--9, 2025, with proton-oxygen (pO), oxygen-oxygen (OO), and neon-neon (NeNe) collisions.  Early experimental results provide strong evidence of quark-gluon plasma (QGP) formation in these small systems.  I review the motivation for the light-ion collisions and the first experimental results, connecting perturbative QCD, hot QCD, and low-energy nuclear structure physics.}

\section{Two Paradigms of Hadron Collisions}

\begin{figure}[b]
  \centerline{\includegraphics[width=0.9\linewidth]{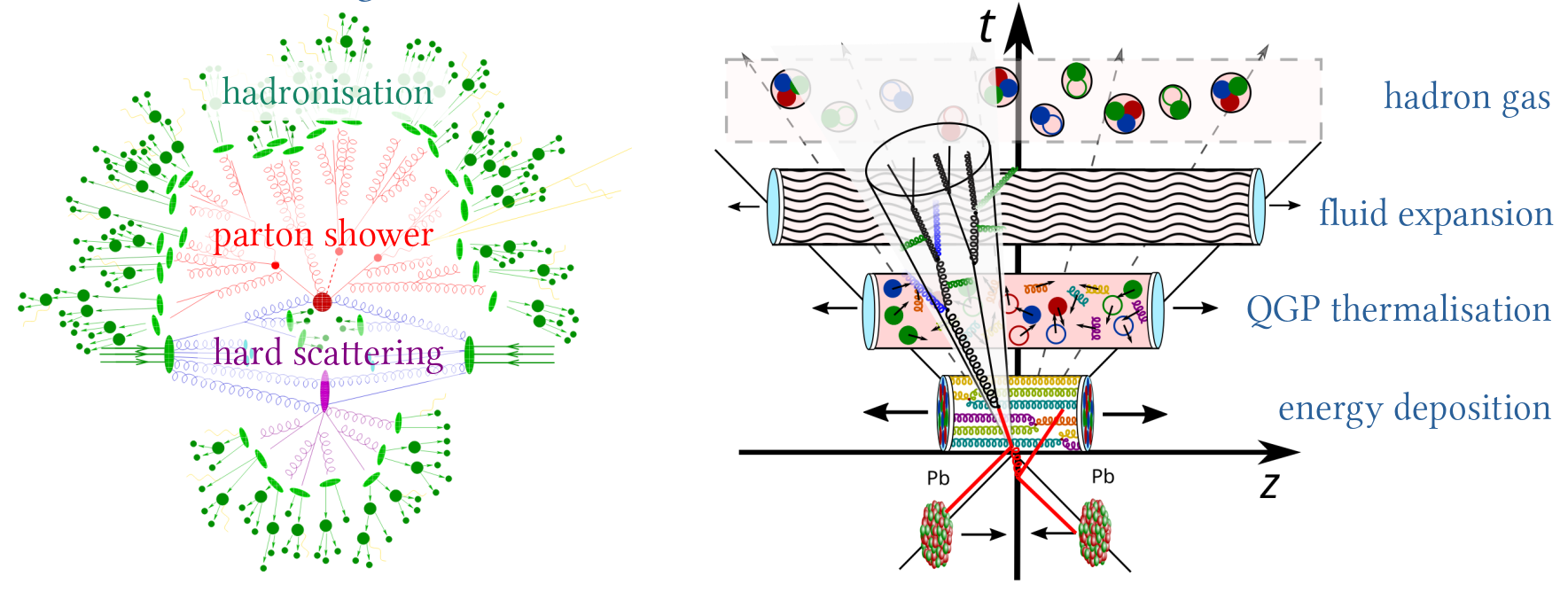}}
  \caption[]{Two pictures of high-energy hadron collisions. (left) Free-streaming final state following hard scattering. Adapted from~\cite{Gleisberg:2008ta}. (right) Thermalisation and hydrodynamic expansion of the QCD medium. Adapted from~\cite{Berges:2020fwq}.}\label{fig:pictures}
\end{figure}

The high-energy hadron collisions at the LHC are described by two starkly contrasting pictures, shown in Fig.~\ref{fig:pictures}. The high-energy event generators
describe hadron collisions as a hard partonic scattering followed by the partonic
shower phase without final-state rescatterings. In the heavy-ion physics
picture, the produced partons thermalise into a quark-gluon plasma (QGP), whose expansion is well described by relativistic viscous hydrodynamics~\cite{Berges:2020fwq,Busza:2018rrf}.

The key evidence for QGP formation is the observation of long-range
two-particle correlations---the elliptic flow~\cite{Ollitrault:1992bk}. The initial nuclear
overlap creates a transversely anisotropic region that extends along
the beam direction. Pressure-gradient-driven expansion of the QGP
converts this spatial anisotropy into a momentum-space anisotropy that is long-range
in pseudorapidity.
The observed momentum anisotropies are quantified by harmonic flow coefficients $v_n$
and are well described by multi-stage hydrodynamic simulations~\cite{Bernhard:2019bmu}.

Surprisingly, the same signatures of elliptic flow have been observed in
high-multiplicity proton-proton collisions already in the first year of LHC running~\cite{CMS:2010ifv}. By now
the collective flow signals have been seen in minimum-bias proton-proton, proton-nucleus, and peripheral nucleus-nucleus collisions,
demonstrating that collectivity
is a generic feature of QCD~\cite{Grosse-Oetringhaus:2024bwr}.

Further evidence for the strongly interacting QCD medium is the observed suppression of
high-momentum hadrons and jets, known as jet
quenching~\cite{Wang:2025lct}. The high-momentum partons interact strongly
with the QGP and shed their energy through medium-induced gluon radiation, see Fig.~\ref{fig:pictures}(right).
The suppression is quantified by
the nuclear modification factor $R_\mathrm{AA}$, defined as the
ratio of the nucleus-nucleus yield to the $pp$ yield scaled by the
number of binary nucleon-nucleon collisions.
In central Pb-Pb
collisions, jets are suppressed by up to 40\%, and
the suppression decreases in more peripheral collisions due to the smaller size of the produced QGP.
In the most peripheral Pb-Pb collisions, the systematic
uncertainties in the number of binary collisions prevent conclusions about energy loss in systems with
just ${\sim}\,10$ participating nucleons. Similarly, energy loss measurements in p-Pb
collisions have been inconclusive~\cite{Grosse-Oetringhaus:2024bwr}.
The
coexistence of collective flow with inconclusive jet-quenching
signatures in these systems has been dubbed the ``small system puzzle.''

\section{Physics Opportunities with Light-Ion Collisions}

\begin{figure}[b]
  \centering
  \includegraphics[width=0.50\linewidth]{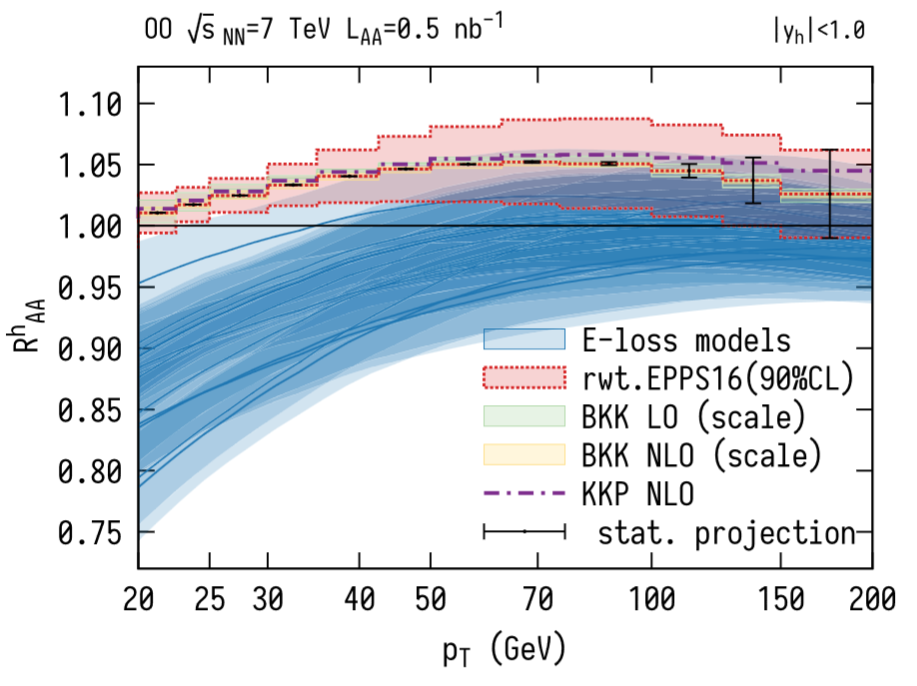}
  \includegraphics[width=0.45\linewidth]{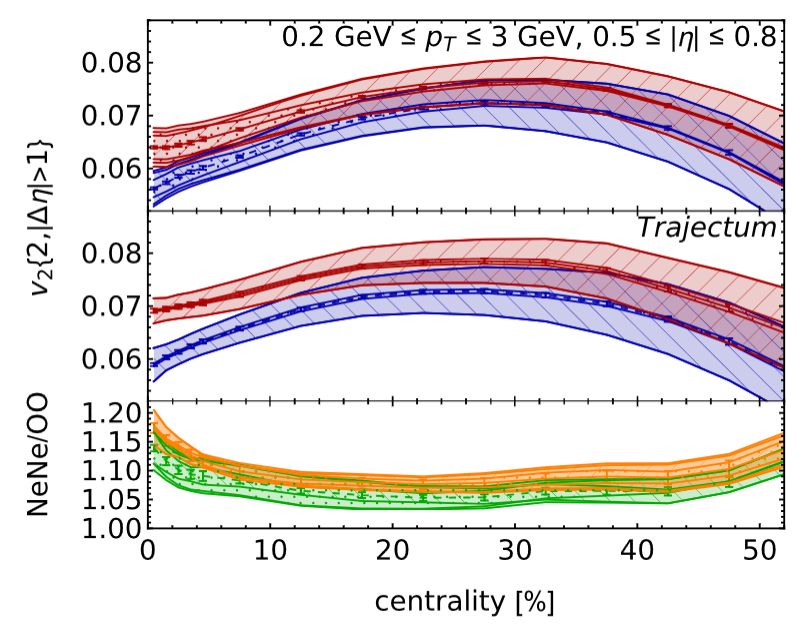}
\caption[]{(left) Charged hadron $R_\mathrm{AA}$: no-quenching baseline and energy loss model predictions for OO collisions. From~\cite{Huss:2020dwe}. (right) Elliptic flow predictions in OO and NeNe collisions. From~\cite{Giacalone:2024luz}.}\label{fig:predictions}
\end{figure}

A new opportunity to address the small system puzzle arose with a short
oxygen programme, originally proposed for cosmic-ray studies, at LHC Run~3~\cite{Citron:2018lsq,Brewer:2021kiv}.
In minimum-bias oxygen-oxygen (OO) collisions, the nuclear modification factor reduces to the ratio of cross-sections
\be
R_\mathrm{OO}^\mathrm{min.\,bias}
  = \frac{1}{A^2}\frac{d\sigma_\mathrm{OO}/dp_T}{d\sigma_{pp}/dp_T}\,,
\ee
with $A=16$, which is free of binary collision modelling. The high-momentum cross-section can be systematically computed using standard pQCD techniques and the
global extractions of nuclear parton distribution functions (nPDFs).

In 2020, we computed the no-quenching baseline of jet and hadron $R_\mathrm{AA}$ in OO collisions~\cite{Huss:2020dwe}. We supplemented these with energy loss model predictions anchored on heavy-ion collision data~\cite{Huss:2020whe}.
We found that the dominant theoretical uncertainty in the no-quenching baseline
is due to nPDFs. As shown in Fig.~\ref{fig:predictions}(left), we identified a potential window of discovery for charged-hadron
$R_\mathrm{AA}$ in the range
$20\,\mathrm{GeV}<p_T<100\,\mathrm{GeV}$, where the energy-loss
signal is well separated from the nPDF uncertainty band.
Recently, we have updated the baseline computations with new nPDF extractions~\cite{Gebhard:2024flv,Mazeliauskas:2025clt,Jonas:2026yoz}.

An unexpected motivation for light-ion runs was the ability of flow observables to probe the two-point density of the ground-state wave function of the colliding nucleus~\cite{Giacalone:2019pca}, which for light nuclei can be computed using state-of-the-art chiral EFT.
While the $^{16}$O nucleus has a compact shape of four
$\alpha$-clusters, the $^{20}$Ne nucleus contains an additional
$\alpha$-cluster that results in a large intrinsic shape
deformation.\footnote{Strictly speaking, we are sensitive to
harmonic modulations of two-body density of the nuclear ground-state wave
function, intuitively illustrated by intrinsic shape deformations~\cite{Blaizot:2025bfu}.}
The hydrodynamic simulations then predicted an enhanced elliptic flow in central NeNe
collisions compared to OO~\cite{Giacalone:2024luz}, see Fig.~\ref{fig:predictions}(right). Crucially, the systematic uncertainties of QGP modelling cancel in the ratio, making this ratio a precision observable of nuclear structure. This motivated an addition of a single fill of neon at the LHC.

The light-ion run took place in July 2025 and was extremely successful: all four
major LHC experiments collected data well above the target
luminosity. In addition to pO ($\sqrt{s_\mathrm{NN}}=9.62\,\mathrm{TeV}$), OO and NeNe ($\sqrt{s_\mathrm{NN}}=5.36\,\mathrm{TeV}$) data in collider mode,
LHCb SMOG2 took NeNe and OH$_2$ data in fixed-target mode. Finally, the LHC data are complemented by OO collision data taken at $\sqrt{s_\mathrm{NN}}=200\,\mathrm{GeV}$ at
RHIC in the 2021 and 2026 campaigns, which provide an additional lever arm in collision energy.

The early experimental results for flow observables are in excellent agreement with
hydrodynamic predictions indicative of QGP formation in light-ion collisions~\cite{ALICE:2025luc,ATLAS:2025nnt,CMS:2025tga}.
The hadron nuclear modification factors show significant suppression compared to $pp$ collisions~\cite{CMS:2025bta,CMS:2026qef}, providing evidence for medium-induced energy loss. However, the residual nPDF uncertainty prevents an unambiguous ${\geq}\,5\sigma$ attribution of this suppression to final-state energy loss. Upcoming analyses of pO data will be crucial for reducing nPDF uncertainties~\cite{Jonas:2026yoz}.

\section{Summary and Outlook}

The successful light-ion collision run at the LHC has been the result of close collaboration between theorists, accelerator specialists, and experimentalists.
The early results show strong evidence that the same type of high-temperature QGP phase created in heavy-ion collisions is also produced in light-ion collisions with
just ${\sim}\,10$ participating nucleons on average. Crucially, this demonstrates that
the hot and dense QCD models trained on heavy-ion data can make precise
quantitative predictions for new collision systems. This increases our confidence
in the understanding of the QGP medium. The first evidence of hadron suppression in OO and NeNe collisions represents an important step towards resolving the ``small system puzzle.'' Nonetheless, how the two pictures of hadron collisions in Fig.~\ref{fig:pictures} can be reconciled remains an open question.
The well-controlled environment of light-ion collisions may prove to be the
crucial testing ground for understanding this transition.

The light-ion run has demonstrated that even short collision campaigns
with new ion species can bring a large physics yield and address novel physics questions.
Motivated by this success, we launched a ``New Ions @ LHC Run 4''
initiative with the aim of documenting science cases for various ion
species and producing a white paper by 2029, in time for the
planning of Run~4~\footnote{See \url{https://indico.cern.ch/event/1655628}}.



\section*{Acknowledgments}

The author is supported by the Deutsche Forschungsgemeinschaft (DFG) through the Emmy Noether Programme (project number~496831614) and the Collaborative
Research Centre SFB\,1225 (ISOQUANT) (project number~27381115).

\section*{References}
\bibliography{mazeliauskas_2}

\end{document}